\documentclass[11pt,preprint]{aastex}
\usepackage{apjfonts}

\def\gax{\mathrel{\raise.3ex\hbox{$>$}\mkern-14mu\lower0.6ex\hbox{$\sim$}}}
\def\lax{\mathrel{\raise.3ex\hbox{$<$}\mkern-14mu\lower0.6ex\hbox{$\sim$}}}
\def\gtorder{\mathrel{\raise.3ex\hbox{$>$}\mkern-14mu
             \lower0.6ex\hbox{$\sim$}}}
\def\ltorder{\mathrel{\raise.3ex\hbox{$<$}\mkern-14mu
             \lower0.6ex\hbox{$\sim$}}}

\begin{document}

\title{Failed Supernovae Explain the Compact Remnant Mass Function}

\author{
   C.~S. Kochanek$^{1,2}$ 
  }

\altaffiltext{1}{Department of Astronomy, The Ohio State University, 140 West 18th Avenue, Columbus OH 43210}
\altaffiltext{2}{Center for Cosmology and AstroParticle Physics, The Ohio State University, 191 W. Woodruff Avenue, Columbus OH 43210}

\begin{abstract}
One explanation for the absence of higher mass red supergiants 
($16.5 M_\odot \ltorder M \ltorder 25M_\odot$)
as the progenitors of Type~IIP supernovae (SNe) is that they die in
failed SNe creating black holes.  Simulations show that such failed
SNe still eject their hydrogen envelopes in a weak transient, 
leaving a black hole with the mass of the star's helium core ($5$-$8M_\odot$).  
{\it Here we show that this naturally explains the typical masses of
observed black holes and the gap between neutron star and black 
hole masses without any fine-tuning of the SN mechanism beyond
having it fail in a mass range where many progenitor models have
density structures that make the explosions more likely to fail.}    
There is no difficulty including this $\sim 20\%$ population of
failed SNe in any accounting of SN types over the progenitor
mass function.  And, other than patience, there is no observational
barrier to either detecting these black hole formation events
or limiting their rates to be well below this prediction.
\end{abstract}

\keywords{stars: evolution -- supergiants -- supernovae:general --
}

\section{Introduction}
\label{sec:introduction}

All massive $M\gtorder 8M_\odot$ stars undergo core collapse, but only some must explode as
core collapse supernovae (ccSNe).  All collapses leading to the formation
of a neutron star must have a ccSNe to eject mass and avoid collapse to
a black hole.  Core collapse can lead to the formation of a black hole either
through a failed SN, where the stalled accretion shock never revives, or
in a successful ccSN where sufficient mass falls back onto the proto-neutron star 
to cause collapse to a black hole.  Little is observationally known about
the balance between these scenarios.  The diffuse supernova neutrino background sets
an upper limit on the failed SN rate at roughly $50$-$75\%$ of the observed SN
rate (\citealt{Lien2010}), and there is some evidence for a mismatch between
massive star formation and SN rates, suggesting a significant failed SN
rate (\citealt{Horiuchi2011}, but see \citealt{Botticella2012}), but
theoretical studies generally favor low rates of failed SNe ($\sim 10\%$ of ccSN rate) at Solar 
metallicity (e.g. \citealt{Woosley2002}).  We know nothing observationally
about the formation of black holes in successful ccSNe, but it is a
relatively common outcome in simulated explosions (e.g. \citealt{Zhang2008},
\citealt{Fryer2012}).

While not directly motivated by understanding the formation of black holes, surveys
attempting to provide a census of the progenitor stars to successful SNe
can reveal the existence of failed SNe.
In particular, there appears to be a deficit of high mass progenitor
stars compared to standard initial mass functions (\citealt{Kochanek2008}, \citealt{Smartt2009}), 
This is best quantified for the deficit of higher mass ($\sim 20M_\odot$) red
supergiant progenitors (\citealt{Smartt2009}), and it is interesting to note that this mass range
also corresponds to stars with internal structures that make it more difficult
for them to explode (e.g. \citealt{Oconnor2011}, \citealt{Ugliano2012}). 
Perhaps this deficit can be explained by observational biases such as
stronger dusty winds around more massive stars (\citealt{Walmswell2012}, but
see \citealt{Kochanek2012}), or by having stars in this mass range evolve away
from being red supergiants before exploding (see the discussion in \citealt{Smartt2009}),
but one simple explanation is that stars in
the mass range $16.5 M_\odot \ltorder M \ltorder 25 M_\odot$ form black
holes without a SN.  The lower
limit is set by the upper mass limit \cite{Smartt2009} found for Type~IIP 
progenitors and $25M_\odot$ is a reasonable estimate for the maximum mass
of stars that undergo core collapse as red supergiants (see the discussion 
in \citealt{Smartt2009}).  This mass range corresponds to $\sim 20\%$
of core collapses.

In principle, neutrino (e.g. \citealt{Abbasi2011}, \citealt{Alexeyev2002}, \citealt{Ikeda2007}, 
or gravitational wave detection (e.g. \citealt{Ott2009}) of a core collapse 
leading to black hole formation, combined with external astronomical observations 
of any resulting transient, would be the cleanest probe of this phenomenon. 
Unfortunately, such observations are only feasible in our Galaxy
and its very nearest neighbors (e.g. \citealt{Ando2005}, \citealt{Scholberg2012}), so 
the event rates are unpleasantly low if the failed SN rate is $\sim 20\%$
of the SN rate.  While any associated visible transient would be more easily
observed in a nearby galaxy, almost all events in the Galaxy would be observable
in the near-IR despite the high extinction in the Galactic plane (see \citealt{Adams2013}).

After pointing out the deficit of higher mass SN progenitors in \cite{Kochanek2008},
we also outlined an approach to identifying black hole formation events without
accompanying SN that was independent of the nature of any intervening transient.
One carries out a ``disappearance'' experiment, monitoring
a large number of evolved massive stars to see if any ``vanish''.  We advocated this
approach because the nature of the phenomenology that occurs between presence
and absence was unclear and little studied.  Some stars have density
structures that would literally allow a direct collapse with almost no
external signature (\citealt{Woosley2012}), others might go through a phase where accretion
onto the the black hole temporally and almost certainly unstably supports the 
remaining stellar envelope, or a weak shock might reach the surface of the
star to create a weak explosive transient (e.g. \citealt{Nadezhin1980}, 
\citealt{Lovegrove2013}, \citealt{Piro2013}).  Our point in \cite{Kochanek2008} was 
that a search for black hole formation in failed SNe could be carried out independent
of whether there was an associated, external transient because the initial and final states 
of the system were well defined.  Moreover, this is the only direct detection method that
is likely to succeed in the near term given supernova rates and the limited
sensitivities of present and future neutrino and gravitational wave detectors.

\begin{figure}[t] 
\centerline{\includegraphics[width=5.5in]{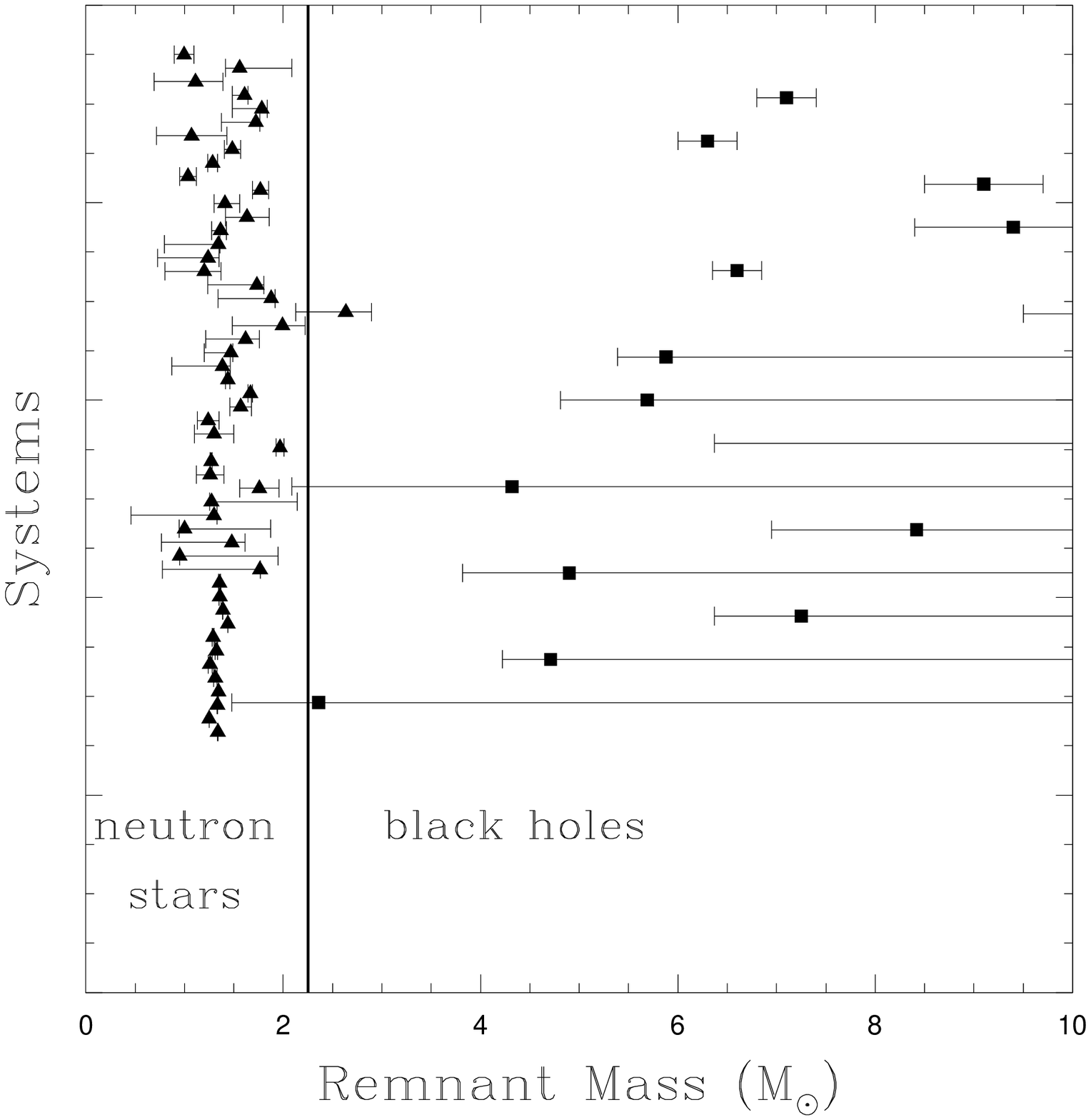}}
\caption{ 
  Observed masses of neutron stars (filled triangles) and black 
  holes (filled squares) from \cite{Ozel2010} and \cite{Ozel2012}.
  The thick solid line at $2.25M_\odot$ roughly marks the maximum
  mass of a neutron star.  The relative numbers of neutron stars
  and black holes cannot be quantitatively compared.
  } 
\label{fig:ozel}
\end{figure}

The last available probe of these physical processes is the mass function
of the resulting neutron stars and black holes.  We show the observed masses
of these remnants in Figure~\ref{fig:ozel} using the summaries from \cite{Ozel2010}
and \cite{Ozel2012}.  While this has many
observational issues because we can only measure masses in binary 
systems, the observed mass function has three striking
features (see, e.g., \citealt{Bailyn1998}, \citealt{Ozel2010}, \citealt{Farr2011}, 
\citealt{Kreidberg2012}, \citealt{Ozel2012}).  
First, neutron star masses have a narrow distribution.  Second,
there is a significant gap between the mass distribution of neutron
stars and black holes.  Third, while very high mass black holes exist,
the typical black hole is far less massive than the stars believed to
have created it.  \cite{Ozel2010}
argue that selection effects due to the requirements for producing accreting
binaries may bias the observed distributions against higher mass black
holes ($M \gtorder 10M_\odot$) but should not be producing a gap between neutron
stars and black holes.

\begin{figure}[t] 
\centerline{\includegraphics[width=5.5in]{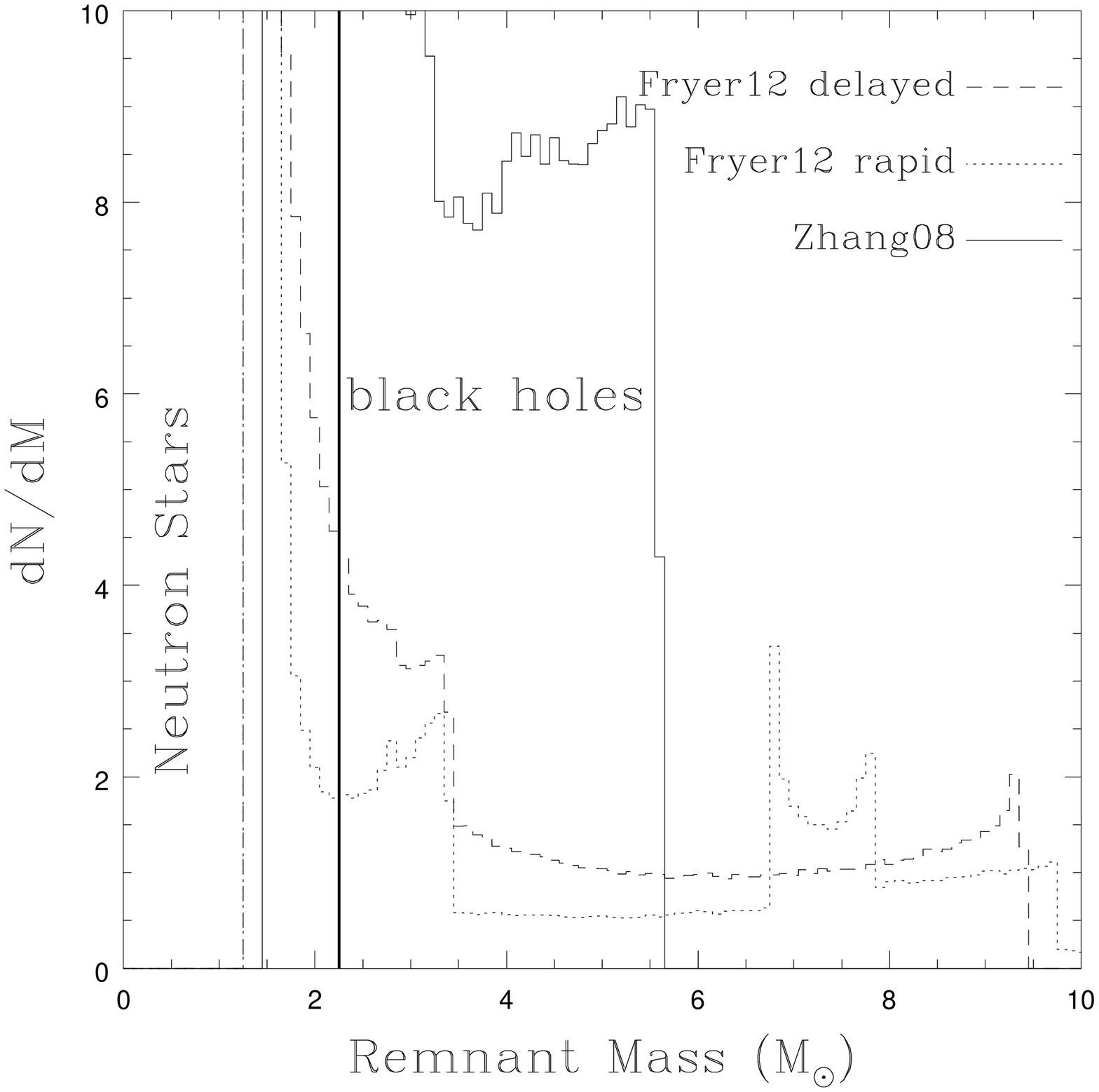}}
\caption{ 
  Remnant mass distributions for the explosion models of \cite{Zhang2008}
  (solid) and \cite{Fryer2012}, where dotted lines show the distribution
  for rapid explosions and dashed lines show the distribution for delayed
  explosions.  The distributions are normalized by the average number
  of remnants between $5M_\odot$ and $10M_\odot$.   The models predict
  distributions dominated by low mass ($<5M_\odot$) black holes and
  lack any clear gap between the masses of neutron stars and black holes. 
  The thick solid line at $2.25M_\odot$ roughly marks the maximum
  mass of a neutron star.  
  } 
\label{fig:old}
\end{figure}

The simplest physical possibility would have been that successful SNe
explosions lead to neutron stars with (roughly) the mass of the iron core 
(less neutrino losses etc.) and failed explosions lead to a
black hole with the mass of the star at death. This would naturally lead 
to a gap in mass, with neutron stars having a mass $\simeq 1.4M_\odot$ and 
black holes having the final stellar mass.   The problem is that almost
all stars at death are more massive than the typical mass of the observed
black holes.  In standard models (e.g. \citealt{Zhang2008}, \citealt{Fryer2012}), 
this forces black holes to be made in the course of successful ccSNe explosions 
so that mass can be ejected.    The explosion is initiated in the
core and the initial mass of the proto-neutron star is similar to
an explosion in which a neutron star is formed, but there is then
significant ``fall back'' of material onto the neutron star leading
to the formation of a black hole.  This allows formation of black holes
less massive than their progenitor stars, but requires fine tuning of
stellar mass loss and explosion energies.  In particular, it is 
difficult to avoid a continuous distribution of remnant masses 
without the observed gap between neutron stars and black holes.
These problems are illustrated in Figure~\ref{fig:old}, where we show
the remnant mass distributions predicted by \cite{Zhang2008} and 
\cite{Fryer2012}. We fully explain the construction of Figure~\ref{fig:old}
in \S2.
The \cite{Zhang2008} models have no gap, while the \cite{Fryer2012}
models partially create one by changing the explosion energetics with
progenitor mass (also see \citealt{Belczysnski2012}).  One practical difference between these models is
that the \cite{Fryer2012} models were constructed in part
to explain the remnant mass distribution while the \cite{Zhang2008}
models were not.  Perhaps the lack of similarity between the observed
(Figure~\ref{fig:ozel}) and model (Figure~\ref{fig:old}) distributions
is purely observational, but perhaps it is also evidence that fall back
is not a good mechanism for explaining the masses of black holes.

{\it In fact, simply accepting the evidence from progenitor studies that red supergiant
stars with masses of $16.5 M_\odot \ltorder M \ltorder 25 M_\odot$ suffer failed supernova explosions 
and form black holes provides a new and very natural explanation for both the
existence of the gap and the typical masses of black holes.}
The key is the observation by \cite{Nadezhin1980} that the envelopes
of red supergiants are so weakly bound that the weakening of the gravitational
potential created by the mass lost in neutrinos during core collapse
is sufficient to unbind the hydrogen envelope of the star.  \cite{Lovegrove2013} 
carried out detailed radiation-hydrodynamic simulations of this mechanism for 
$15M_\odot$ and $25M_\odot$ red supergiants, finding that the adjustment of the 
envelope to the neutrino mass loss can trigger a weak shock that unbinds the 
envelope as \cite{Nadezhin1980} predicted. 
The result is a low luminosity ($\sim 10^6 L_\odot$), cool ($\simeq 3000$~K) 
transient lasting roughly one year and largely powered by the recombination energy 
of the envelope. \cite{Piro2013} notes that there is also a shock
break out pulse which is $10$-$30$ times brighter, hotter ($\simeq 10^4$~K)
and lasting roughly a week.  

The key point for the remnant mass distribution is that the natural
mass scale of the resulting black hole is the mass of the helium core
of the progenitor star, thereby reproducing both the gap between
neutron star and black hole masses and the characteristic minimum
masses of black holes with no need for fine tuning stellar mass loss,
the explosion mechanism, the amount of fall back, or binary evolution.  In \S2
we show some simple models of mass functions based on this
idea and how to fit a significant population of failed SNe into
the overall accounting for the deaths of massive stars. 
In \S3 we discuss some additional implications and strategies
for identifying these events.  

\section{Results}

We generated Figure~\ref{fig:old} using the following assumptions.  We
drew the progenitor masses from a Salpeter IMF over the mass range $8.5M_\odot < M < 100M_\odot$,
although the upper mass limit is quantitatively unimportant.  We then 
assigned masses either by interpolating over the \cite{Zhang2008} models
or using the analytic approximations in \cite{Fryer2012}. We used the
Solar metallicity SA model from \cite{Zhang2008}, corresponding to an
energy of $1.2$~Bethe and a piston located at a fixed entropy per particle
of $S/k=4$.  Above and below the tabulated range from $12$ to $100M_\odot$
we simply used the results for the appropriate limiting mass.  For the 
\cite{Fryer2012} models we assigned $1.4M_\odot$ remnant masses to 
progenitors with $M<11M_\odot$.  As noted in \S1, the results predict 
a continuum of black hole and neutron star masses and lack a clear peak
in the observed mass range of black holes.  When comparing Figures~\ref{fig:ozel}
and \ref{fig:old} one should not be comparing the relative numbers of black
holes and neutron stars, but only the two classes separately.  In \cite{Pejcha2012}
we modeled the masses of binary neutron stars from \cite{Ozel2012} and found
that they strongly disfavored models in which there was any fall back mass,
while all the models used in Figure~\ref{fig:old} must include fall back because
it is the only way they can produce any low mass black holes.

\begin{figure}[t]
\centerline{\includegraphics[width=5.5in]{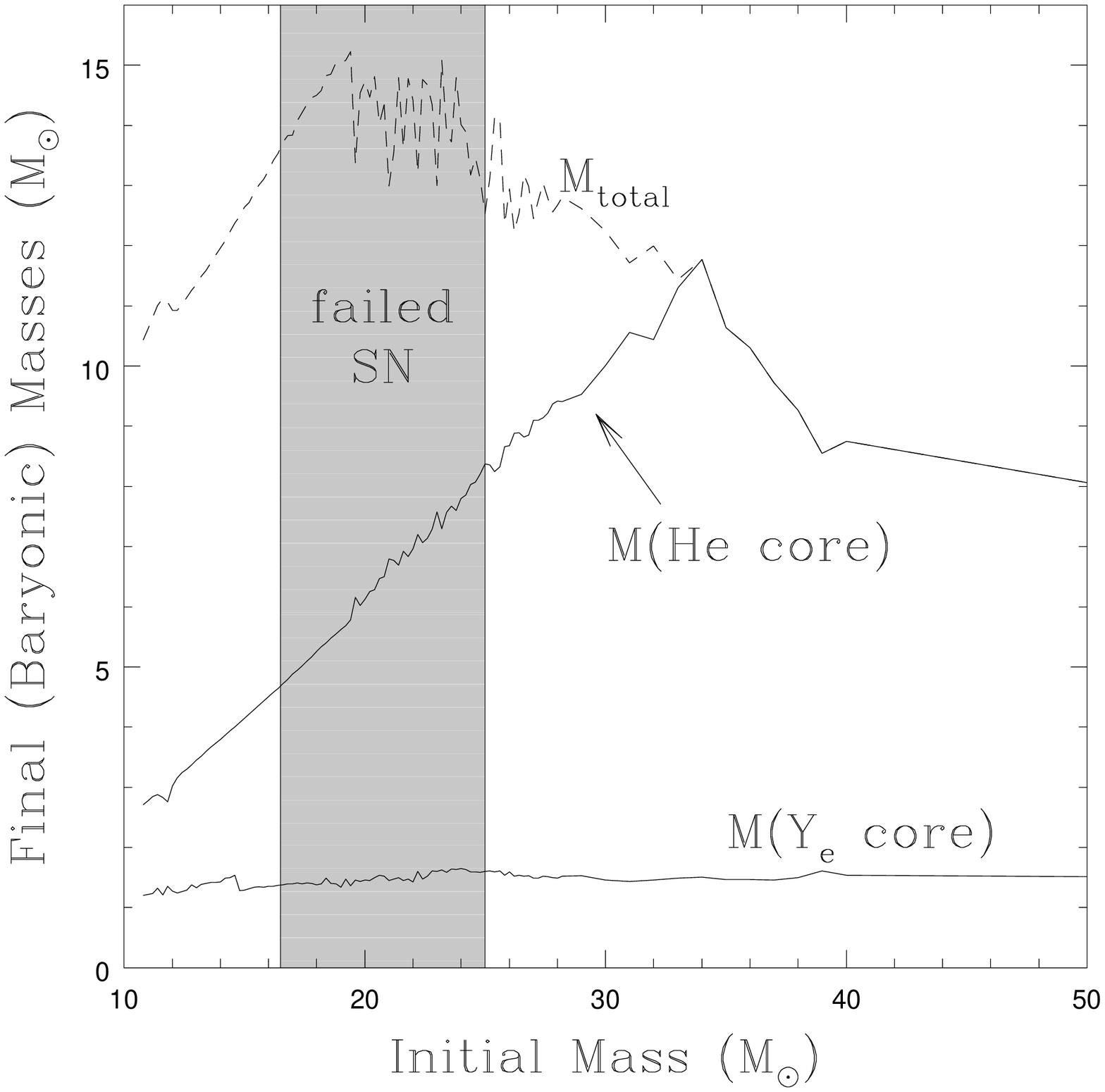}}
\caption{
  Pre-supernova structure of massive stars from \cite{Woosley2002}.
  The dashed line shows the final mass of the star.
  The solid lines show the mass of the $Y_e$ core, which will 
  roughly correspond to the mass of any resulting neutron star,
  and the mass of the helium core.  Note the enormous mass loss
  associated with the higher mass stars.  The shaded region
  encompass the mass range from $16.5M_\odot$ to $25M_\odot$
  that we associate with failed supernovae.
  }
\label{fig:pmass}
\end{figure}

In the \cite{Nadezhin1980} mechanism, as confirmed by the simulations of
\cite{Lovegrove2013}, the remnant mass from the failed SN of a red supergiant
is set by the mass of the helium core of the star.  There may
be some hydrogen fall back contribution, but in the \cite{Lovegrove2013}
simulations it is small.  Figure~\ref{fig:pmass} shows the pre-SN mass,
helium core mass, and the mass of the $Y_e$ core for the pre-SN stellar 
models of \cite{Woosley2002}.  The $Y_e$ core mass is a good proxy for
neutron star masses in the absence of fall back and is defined by the
point in the core where there is a significant jump in the electron 
abundance $Y_e$.
In \cite{Pejcha2012} we found that the mass of the 
$Y_e$ core with no fall back was one of the better models
for the mass distribution of binary neutron stars. 
In this model sequence, mass loss becomes increasingly important 
above $20M_\odot$.  Stars in the mass range corresponding 
to the missing red supergiant progenitors have helium core
masses of between $5M_\odot$ and $8M_\odot$, almost exactly
corresponding to the observed mass range of black holes. 

\begin{figure}[t]
\centerline{\includegraphics[width=5.5in]{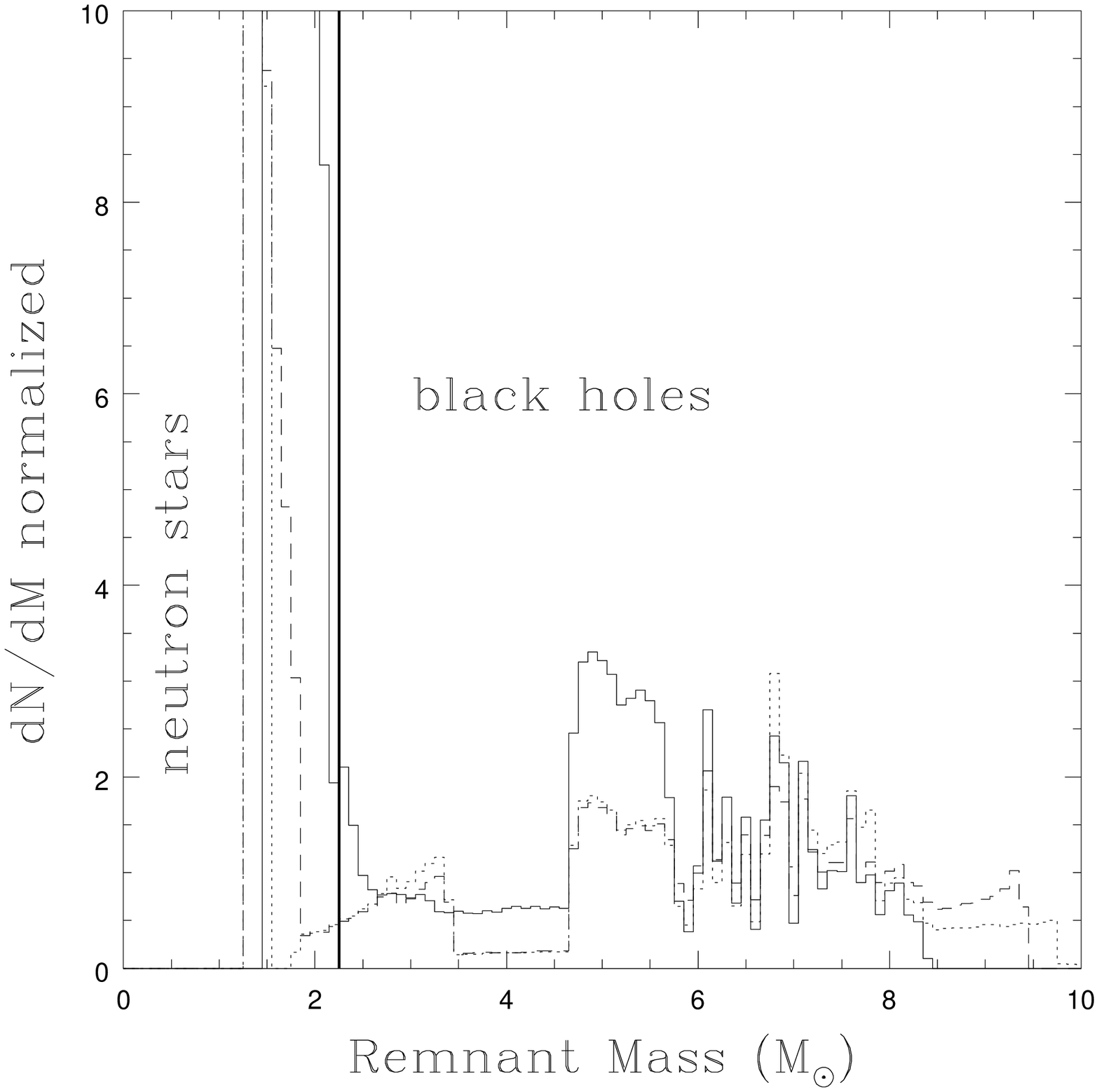}}
\caption{
  Remnant mass distributions if the core collapse of stars in the mass
  range from $16.5M_\odot$ to $25M_\odot$ leads to the formation of 
  black holes with the helium core mass from the models of 
  \cite{Woosley2002} (Figure~\ref{fig:pmass}). Outside this mass
  range we continue to use the results from \cite{Zhang2008} (solid)
  and \cite{Fryer2012},  where dotted lines show the distribution
  for rapid explosions and dashed lines show the distribution for delayed
  explosions.  The distributions are normalized by the average number
  of remnants between $5M_\odot$ and $10M_\odot$.  The black hole
  mass distributions now have a distinct peak in the observed mass
  range and far fewer low mass black holes.  These low mass black
  holes could be completely eliminated by adopting explosion models
  with no fall back.  
  The thick solid line at $2.25M_\odot$ roughly marks the maximum
  mass of a neutron star.
  }
\label{fig:new}
\end{figure}

We can now examine the remnant mass distribution if the more massive red
supergiants undergo failed SN but eject their hydrogen envelopes based on
the \cite{Nadezhin1980} mechanism. 
Figure~\ref{fig:new} shows the remnant mass distributions 
after we replace the remnant masses from the underlying models
for the mass range $16.5M_\odot < M < 25M_\odot$
with the helium core mass shown in Figure~\ref{fig:pmass}.
For all three cases, there is now a far more distinct peak at the observed
masses of black holes, and a greatly reduced production of unobserved low
mass black holes. The effect is most dramatic for the \cite{Zhang2008} model
where the fraction of black holes (remnant masses $>2M_\odot$) with masses
between $5$ and $10M_\odot$ rises from 
6\% to 46\%.  Even for the \cite{Fryer2012} models, where the parameters
were in part tuned to better reproduce the mass function of binary black holes,
the fractions rise from $\sim 40\%$ to $\sim 70\%$.  

\cite{Smith2011} argue that adding such a population of
failed SNe is difficult to reconcile with attempts to
distribute SN types over the IMF.    The essence of the
argument is that the Type~IIP fraction of $(48\pm6)\%$ 
found by \cite{Li2011} is so low that the proposed
failed SN mass range of $16.5 M_\odot \ltorder M \ltorder 25 M_\odot$
needs to produce non-Type~IIP SNe in order to match the 
observed SN type fractions.  For example, in
the absence of binaries, assigning the mass range 
from $M_0=9.4M_\odot$ to $M_1=15.3M_\odot$ to producing Type~IIP
SNe, and $M>M_1$ to producing non-Type~IIP SNe has
a 48\% Type~IIP fraction and is consistent ($\chi^2=1.5$ for
1~dof) with the estimates of $M_0=8.5_{-1.5}^{+1.0}M_\odot$ and 
$M_1=(16.5\pm1.5)M_\odot$ by \cite{Smartt2009}. However,
this leaves no room to allot a mass range containing 
$\sim 20\%$ of progenitors to failed SNe.

Rather than being evidence against failed SNe, this is really evidence
for the importance of binaries, which was discussed by \cite{Smith2011}
in other contexts.  Since we lack a fully quantitative understanding
of either mass loss by individual stars or mass transfer in 
binaries, we only consider three classes of objects. There
are Type~IIP SN, which we attribute to lower mass stars 
($M_0 \simeq 8.5M_\odot < M < M_1 \simeq 16.5M_\odot$). These 
limits are chosen to match the mass range found to be associated with Type~IIP
SNe by \cite{Smartt2009}.   
Non-IIP SN are a combination of high mass stars ($M>M_2 \simeq 25M_\odot$)
that have lost mass due to either stellar evolution or binary
mass transfer and a fraction $b$ of the IIP mass range where
binary interactions lead to a non-IIP SN.  The mass scale $M_2 \simeq 25M_\odot$
is roughly the highest mass at which stars both undergo core collapse
as red supergiants and the range of progenitor masses which can be more 
difficult to explode (e.g.  \citealt{Oconnor2011} and \citealt{Ugliano2012}).  Finally, the mass
range $M_1 < M < M_2$ leads to a failed SN.  The interacting
binary fraction is expected to be very high.  For example,
\cite{Sana2012} estimate that for O stars
at birth, roughly 30\% are effectively single, 24\% merge, 33\%
undergo some envelope stripping and 14\% have some accretion, 
which means it is perfectly plausible that $b=30-50\%$ of stars in the
IIP mass range become non-IIP SNe due to interactions.  In the
IIP mass range, the donor star explodes as a non-IIP SN because
of the mass transfer, and the recipient can explode as a non-IIP SN
because it now evolves as a more massive star with greater mass loss.
We discuss the potential effects of this binary fraction on the
phenomenology of the failed SNe in \S3.

\begin{figure}[t]
\centerline{\includegraphics[width=5.5in]{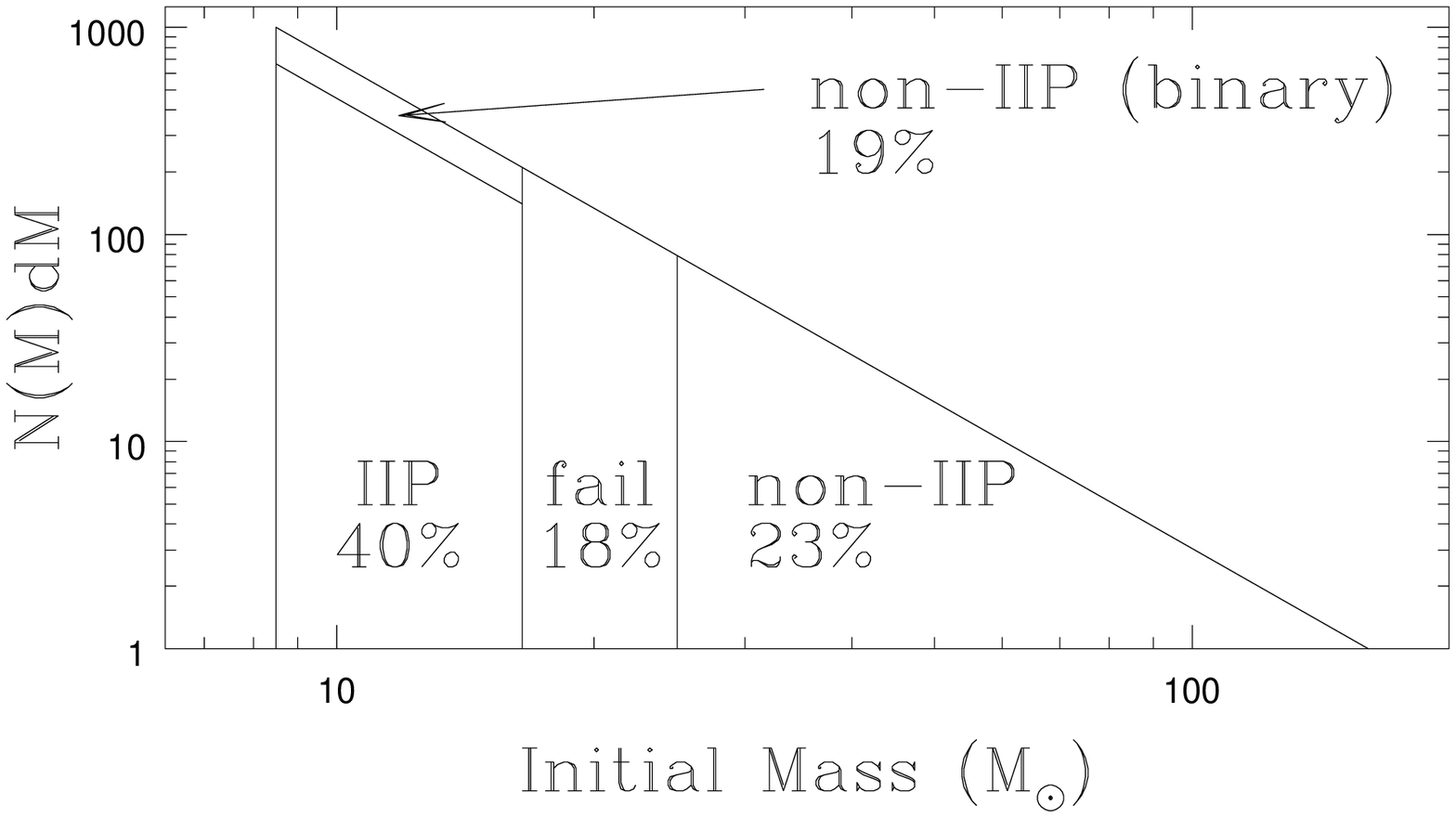}}
\caption{
  Mass ranges implied for SN types following \cite{Smith2011}
  but including a failed SN component.  The fractions represent
  the fraction of all core collapses, so while the fraction of
  core collapses leading to Type~IIP (non-IIP) SNe is 40\% (42\%), 
  the fraction of SNe that are Type~IIP (non-IIP) is 48\% (52\%).  
  Similarly, 18\% of core collapses fail, but the failed SN rate
  is 22\% of the SN rate.  The mass loss making the high mass
  stars have non-IIP SNe can be an arbitrary mixture of binary
  and single star processes.  The interacting binary fraction for this model
  is $b=33\%$, and we discuss how this may affect the phenomenology
  of the failed SNe in \S3.
  }
\label{fig:smith}
\end{figure}

Under these assumptions, we can simply solve for the binary
fraction required to leave a 48\% Type~IIP fraction for
successful SNe.  For $M_0=8.5M_\odot$, $M_1=16.5M_\odot$
and $M_2=25M_\odot$ this results in a binary fraction of
$b=0.33$ which is quite reasonable.  
Raising the upper
mass limit for failed SNe to $M_2=30M_\odot$ only requires
raising the binary fraction to $b=0.37$.  
For $M_2=25M_\odot$ and the higher
Type~IIP fraction of 59\% found by \cite{Smartt2009} 
the binary fraction need only be $b=0.18$.  
Given all the
other uncertainties, the main point is simply that given
a reasonable fraction of binary-induced SN type transformations
(from IIP to not IIP),
there is no difficulty accommodating a significant rate of
failed SNe in an accounting of SN types over the initial
mass function.  Figure~\ref{fig:smith} illustrates this graphically
following the similar figures by \cite{Smith2011}.

\section{Discussion}

Assuming that high mass red supergiants die as failed SNe naturally
solves two observational puzzles: (1) the failure to find progenitors
in this mass range (e.g. \citealt{Kochanek2008}, \citealt{Smartt2009}),
and (2) the peculiar mass distribution of compact
remnants (e.g., \citealt{Bailyn1998}, \citealt{Ozel2010}, \citealt{Farr2011},
\citealt{Ozel2012}).  
While we lack a fully predictive theory of core collapse
events, it is also true that many stellar models in this mass range have 
density structures that render them more difficult to explode 
(e.g. \citealt{Oconnor2011}, \citealt{Ugliano2012}).  

Models of successful SNe can form black holes with masses of $5$ to $10M_\odot$
by carefully tuning the stellar mass loss and the explosion energetics to 
achieve the correct amount of mass fall back.  We used the examples of
\cite{Zhang2008}, which poorly matches the observed black hole mass 
distribution, and \cite{Fryer2012}, which does better.  However, if we simply
excise the mass range from $16.5 M_\odot$ to $25 M_\odot$ from these
models and instead make them failed SNe in which the hydrogen envelope
is ejected by the \cite{Nadezhin1980} mechanism, then we naturally 
create large numbers of black holes in the observed mass range.  The
mass scale comes naturally from the mechanism because only the weakly bound
hydrogen envelope can be ejected.  The black hole mass scale is simply the 
mass of the helium cores of the stars ending their lives as failed SNe. 
Since there is no longer any need to tune the amount of fall back
mass to produce low mass black holes in successful SNe, we could now
allow all successful SNe to form neutron stars with no fall back, 
which would eliminate the residual problem that other progenitor mass
ranges also produce low mass black holes in the \cite{Zhang2008} or
\cite{Fryer2012} models.  Having no fall back would also better match
the neutron star mass distribution (\citealt{Pejcha2012}). Similarly, one could add
a population of failed SN associated with more compact, higher mass 
stars where the final mass would be that of the star at the time
of explosion. The key point for models of the remnant mass distribution
is that the \cite{Nadezhin1980} mechanism provides a means of producing $\sim 7M_\odot$ 
black holes that does not depend on either mass loss or the fall back
mechanism.  

If we build a simple model where stars from $8.5M_\odot$ to $16.5M_\odot$
produce Type~IIP SNe unless modified by binary interactions, stars from
$16.5M_\odot$ to $25M_\odot$ produce failed SNe, and higher mass stars
produce non-IIP SNe, then it is easy to match the \cite{Li2011} SN
type fractions if roughly $1/3$ of potential IIP progenitors are 
modified by binary interactions and become non-IIP SNe.  In such a 
model, 18\% of core collapses fail, corresponding to a failed SN
rate that is $f=22\%$ of the SN rate.  Using the higher IIP fraction
found by \cite{Smartt2009} allows a lower binary modification fraction.
Invoking such a fraction of binary interactions is consistent with
the estimates of \cite{Sana2012} from the statistics of O star binaries. 

If we translate the luminosities and temperatures estimated by 
\cite{Lovegrove2013} and \cite{Piro2013} into magnitudes assuming
black body spectra, the absolute visual magnitude of the 
transient and the shock break out pulses are roughly 
$M_V \simeq -8.7$~mag ($M_I \simeq -11.2$) and 
$M_V \simeq -13.5$~mag ($M_I \simeq -13.7$), 
respectively (\citealt{Adams2013}).  The transient is significantly
brighter at longer wavelengths because of its low temperature.
If we assume that the typical Type~IIP magnitude is $M_{IIP}=-16$~mag,
then a field survey finding $N_{IIP}$ Type~IIP SNe should find
\begin{equation}
    N_{Ned} =  f N_{IIP} 10^{0.6(M_{IIP}-M_{Ned})}
\end{equation}
\cite{Nadezhin1980} 
events, where the magnitude dependence
is created by the (Euclidean) dependence of the survey volume on
transient luminosity.  For the extended, cool transient, the 
expected numbers are very low, $\sim 10^{-3}$ to $10^{-4}f N_{IIP}$ depending
on the survey band pass.  This phase of the transient also does
not stand out significantly from other slow variations in high
luminosity stars.  Hence, as suggested by \cite{Piro2013}, 
normal SN surveys should focus on the shock break out pulse from
these events where the expected number would be $\sim 0.03 f N_{IIP}$.  
Such a search will require cadences closer to daily than weekly in 
order to sample the transient well-enough to have confidence in the 
detection and to motivate a search for the longer duration, fainter 
transient with larger telescopes.  Fortunately, the break out 
peak and duration appear to occupy a region of transient space
without significant, known backgrounds, as they should be 
significantly more luminous than classical novae of the same
duration.  Achieving a 90\% confidence limit of $f<0.1$ requires
a survey where the expectation value is 2.3 events, so surveys
containing $10^3$ Type~IIP SNe that could have detected the break
out peaks of these transients at high efficiency will begin to
provide strong constraints on the existence of this mechanism. 
As presently designed, however, most field surveys for SN have cadences
that will make it difficult to achieve a high efficiency for detection of 
these transients since they will sample the events poorly
(e.g. cadences of 5, 7 and 3~days for PTF, Pan-STARRS1 and LSST, 
respectively, see \cite{Rau2009} for a summary of surveys).

A targeted survey focused on nearby galaxies, such as our more general search
for failed supernovae (\citealt{Kochanek2008}), has little prospect of detecting the 
break out pulse because of its low cadence ($>$ monthly), but would
have no difficulty following the longer transient since it was 
already designed to search for the disappearance of far less 
luminous stars.  Here the rate is limited simply by the rarity
of the SNe, since $N_{IIP} \simeq 1$/year for local galaxies, 
leading to an expected rate of $\simeq f$/year for this class of 
failed supernovae.  Thus, achieving a 90\% confidence limit that
$f<0.1$ requires two decades of monitoring nearby galaxies which 
is painfully long but entirely feasible.  Of course if $f\simeq 0.2$, the
probability of finding such an event in a decade is quite
high (86\%) and it is only for these nearby events that we are 
guaranteed to be able to convincingly say that the progenitor star
has vanished.  

Finally, we should not expect all these events to have the luminous
counterparts predicted by \cite{Lovegrove2013}.  First, as noted by
\cite{Lovegrove2013}, the neutrino mass loss may not be large enough
to trigger envelope ejection in all cases.  Second, like the stars which
eventually have successful SNe, a significant fraction of the stars 
that will become failed SNe will be in binaries and will have part
or all of their hydrogen envelopes stripped before death.  To the 
extent that the interior structure that will lead to a failed SNe
is not significantly altered by the mass loss, these stars will
still end as failed SNe. If there is remaining
hydrogen and it is still in an extended, low binding energy envelope,
then we would still expect a transient associated with core collapse
but it would be weaker.  In some senses, these would be the failed
SN equivalents of Type IIL or IIb SNe.  If the mass loss leads the
envelope to collapse or if all the hydrogen is stripped, then there
would likely be no luminous transient.  Since we are unable to
predict the outcome of core collapse from first principles, it is
difficult to address these scenarios quantitatively.  However, if
$ \sim 1/3$ of stars which would otherwise become IIP SNe do not
do so because of binary interactions, we might expect a similar
fraction of failed SNe to be modified.  

\acknowledgements

I would like to thank J. Beacom, A. Gould, T. Piro, K. Stanek
and T. Thompson for discussions or comments and F. {\"O}zel for supplying
the data used in Figure~\ref{fig:ozel}.

\end{document}